\documentclass[apl,twocolumn,showpacs,superscriptaddress,letterpaper]{revtex4-2}
\usepackage{graphicx,amsmath,amsfonts}
\usepackage{tabularx}
\usepackage{bm}
\usepackage{mathrsfs}
\usepackage{appendix}
\usepackage{bbm}
\usepackage{color}
\def \vk {{\bf k}}
\def \vS {{\bf S}}

\def \mb {\mu_{\rm B}}

\def \vn {{\bf n}}
\def \vz {{\bf z}}

\def \vr {{\bf r}}
\def \vR {{\bf R}}
\def \vp {{\bf p}}

\def \kkon {{{\bf k}^*_1}}
\def \kktw {{{\bf k}^*_2}}

\def \lk {{\hat{l}_{z\vk }}}

\def \kov {\overline{k}}

\begin{document}

\title{{Exact Results for the Orbital Angular Momentum of Magnons on Honeycomb Lattices}}

\author{Randy S. Fishman}
\email{corresponding author:  fishmanrs@ornl.gov}

\affiliation{Materials Science and Technology Division, Oak Ridge National Laboratory, Oak Ridge, Tennessee 37831, USA}
\author{Lucas Lindsay}
\affiliation{Materials Science and Technology Division, Oak Ridge National Laboratory, Oak Ridge, Tennessee 37831, USA}
\author{Satoshi Okamoto}
\affiliation{Materials Science and Technology Division, Oak Ridge National Laboratory, Oak Ridge, Tennessee 37831, USA}

\date{\today}

\begin{abstract}
	We obtain exact results for the orbital angular momentum (OAM) of magnons at the high symmetry points of
	ferromagnetic (FM) and antiferromagnetic (AF) honeycomb lattices in the presence of Dzyallonshinskii-Moriya (DM) interactions.  For the FM honeycomb lattice in the
	absence of DM interactions, the values of the 
	OAM at the corners of the Brillouin zone (BZ) ($\kkon=(0,2\sqrt{3}/9)2\pi/a $, $\kktw=(1/3,\sqrt{3}/9)2\pi/a ,\ldots $) are alternately $\pm 3\hbar /16$ for both magnon bands.  The presence of DM 
	interactions dramatically changes those values by breaking the degeneracy of the two magnon bands.  The OAM values are alternately $3\hbar /8$ and 0 for the lower magnon band and 
	$-3\hbar /8$ and 0 for the upper magnon band.
	For the AF honeycomb lattice, the values of the OAM at the corners of the BZ are $\mp (3\hbar /16)\kappa $ on one of the degenerate magnon bands and $\pm (3\hbar /8) (1+\kappa /2)$ on the other, where 
	$\kappa $ measures the anisotropy and the result is independent of the DM interaction.	
\end{abstract}


\maketitle

\section{Introduction}

The conversion of spin into orbital angular momentum and back has been of great interest since the 
early measurements of Einstein and deHass \cite{Einstein15} and Barnett \cite{Barnett15}.  One recently posed question
\cite{Mat11a, Mat11b, Neumann20, Streib21} is whether spin excitations,
commonly called magnons, can have both spin and orbital angular momentum (OAM).  While the magnon produced by a ferromagnet (FM) with moments up
has spin ${\cal S} = - \hbar $, its OAM  
${\cal L}$ is not known.  The observation of the OAM associated with magnons would have wide-ranging physical ramifications.
Since the total angular momentum ${\cal J}={\cal S}+{\cal L}$ is conserved \cite{Tsukernik66, Garmatyuk68}, 
magnons with ${\cal S}$ parallel to ${\cal L}$ and those with ${\cal S}$
antiparallel to ${\cal L}$ will behave differently under collisions and in external fields.  Consequently, memory storage devices may be able to utilize the OAM of 
magnons in the developing field of magnonics \cite{Karen14}.  It is likely that the magnon Hall \cite{Fuji09, Katsura10, Onose10} and Nernst \cite{Cheng16} effects
will sensitively depend
on the OAM of the magnons.  Another fascinating possibility is that magnons with OAM can
be created and controlled by the interaction with photons \cite{Marrucci06} and phonons \cite{Zhang14, Hamada18} that also carry OAM.

In a recent paper \cite{fishun}, two of us demonstrated that the magnons of collinear magnets may exhibit OAM even for simple collinear magnets.
Surprisingly, the OAM of magnons on FM and antiferromagnetic (AF) zig-zag and honeycomb lattices in two dimensions depends on the network 
of exchange interactions and not on spin-orbit (SO) coupling, as was proposed in Ref.\,\cite{Neumann20}.   Rather, we found that the confinement of  
magnons can generate an effective OAM by disrupting the locally symmetric environment at each lattice site.  
In that sense, our previous efforts followed earlier studies of the OAM of magnons in confined 
spherical 
\cite{Haigh16, Sharma17, Osada18} and cylindrical \cite{Jiang20, Lee22, Kamenetskii21} geometries.  At least formally, our results also reduced to the well-known 
expressions of Matsumoto and Murakami \cite{Mat11a, Mat11b} for the self-rotation of magnon packets in the case of FM systems at low 
excitation energies.

Our previous work was based almost entirely on 
numerical simulations.  In this paper, we obtain
exact results for the OAM of magnons on FM and AF honeycomb lattices at the corners $\vk^*$ of the Brillouin zone (BZ) in reciprocal space
(see Fig.\,2) in the presence of Dzyalloshinzki-Moriya (DM) interactions, which are induced by SO interactions.
In addition to providing checks on our numerics, this work also provides deeper insight into the profound and unexpected effects of the DM interaction on the 
OAM of honeycomb lattices.  For a FM honeycomb lattice, we find that DM interactions can dramatically change the amplitude of the OAM near the BZ corners.

Section II summarizes the OAM formalism for the magnons of FM and AF honeycomb lattices.  Section III presents our results for the FM honeycomb lattice
in the absence of DM interactions.
We then add DM interactions in Section IV.  Section V treats the AF honeycomb
lattice.  Finally, Section VI contains a conclusion.

\section{OAM Formalism}

As sketched in Fig.\,1, honeycombs are non-Bravais lattices
with two unique sites 1 and 2.
The total magnetization ${\bf M}_i$ at site $i$ can be written in terms of the dynamical magnetization $\boldsymbol{\mu}_i$ as
\begin{equation}
{\bf{M}}_i=\boldsymbol{\mu }_i +\vn_i \sqrt{M_0^2 - \mu_i^2},
\end{equation}
where the static magnetization ${\bf{M}}_{0i} = 2\mb S\,\vn_i $ lies along $\vn_i=\pm \vz$ and $\boldsymbol{\mu}_i\cdot \vn_i =0$. 
The ``pseudo-momentum" ${\bf p}_i$ is derived \cite{fishun}
from the energy-momentum tensor $T_{\alpha \beta }$ \cite{Landau60} for a model spin Lagrangian
\begin{equation}
L=\frac{1}{4\mb M_0}\sum_{i=1}^N \bigl( \dot{\boldsymbol{\mu}}_i\times \vn_i \bigr) \cdot \boldsymbol{\mu}_i - H_2,
\end{equation}
where $H_2$ is obtained by expanding the Hamiltonian 
\begin{eqnarray}
H&=& -\sum_{i=1}^N J_{\alpha \beta } \, M_{i\alpha }M_{i+1, \beta } -\sum_{i=1}^N {\bf M}_i \cdot {\bf h}_i \nonumber \\
&-&\frac{K }{4\mb^2}\sum_{i=1}^{N} M_{iz}^2
- \frac{1}{8\pi }\sum_{i=1}^N h_i^2,
\end{eqnarray}
to second order in $\mu_i $. 
Here, the exchange $J_{\alpha \beta }$ includes possible antisymmetric terms like the DM interaction, the magnetic dipole
field ${\bf h}_i$ satisfies $\nabla \times {\bf h}_i=0$ and $\nabla \cdot {\bf h}_i=-4\pi \nabla \cdot {\bf M}_i$,
and $K$ is the easy-axis anisotropy along $\vz $.  It is straightforward to show \cite{Tsukernik66, Garmatyuk68} that 
the Hamiltonian equations of motion for $\boldsymbol{\mu}_i$ can also be obtained from the Euler-Lagrange equations for $L$.
Consequently,
\begin{eqnarray}
\label{mom}
 p_{i\alpha } &=& -\frac{\partial \boldsymbol{\mu}_i }{\partial x_{\alpha }}\cdot \frac{\partial L}{\partial \dot{\boldsymbol{\mu}_i}}\nonumber \\
&=& - \frac{1}{4\mb M_0} \bigl( \vn_i \times \boldsymbol{\mu }_i\bigr)\cdot \frac{\partial \boldsymbol{\mu}_i}{\partial x_{\alpha }}.
\end{eqnarray}
is the momentum at site $i$

Within linear spin-wave theory, the $1/S$ quantization conditions
\begin{equation}
{\mu}_i^+=\mu_{ix}n_{iz}+i\mu_{iy}=2\mb \sqrt{2S\hbar }\,a_i,
\end{equation}
\begin{equation}
{\mu}_i^-=\mu_{ix}n_{iz}-i\mu_{iy}=2\mb \sqrt{2S\hbar }\,a_i^{\dagger }
\end{equation}
are formulated in terms of local Boson operators $a_i$ and $a_i^{\dagger}$ that satisfy the real-space commutation relations
$[a_i,a_j^{\dagger }]=\delta_{ij}$ and $[a_i,a_j]=0$.  Transforming to momentum space with $r$ and $s$ denoting positions within the magnetic unit cell,
we define Boson operators $a_{\vk }^{(r)}$ and $a_{\vk}^{(r)\dagger }$ satisfying the 
commutation relations $[a_{\vk }^{(r)}, a_{\vk'}^{(s)\dagger }]=\delta_{rs}\,\delta_{\vk,\vk'}$
and $[a_{\vk }^{(r)}, a_{\vk'}^{(s)}]=0$.  We may then write the OAM along $\vz $ as \cite{fishun}
\begin{eqnarray}
{\cal L}_z &= &\sum_{i=1}^N (\vR_i \times \vp_i )\cdot \vz \nonumber \\
&=& \frac{1}{8\mb M_0}\sum_{i=1}^N \Bigl\{ {\mu}_i^+  \,\hat{l}_{zi} \, {\mu}_i^- -{\mu}_i^- \, \hat{l}_{zi} \, {\mu}_i^+ \Bigr\}\nonumber \\
&=&\frac{\hbar}{2}\sum_{r=1}^2 \sum_{\vk } \Bigl\{ a_{\vk}^{(r)}\,\lk \, a_{\vk }^{(r)\dagger }-a_{\vk }^{(r)\dagger }\, \lk \, a_{\vk }^{(r)} \Bigr\},
\end{eqnarray}
where
\begin{equation}
\label{lzr}
\hat{l}_{zi} = -i \bigg( x_i \frac{\partial}{\partial y_i} - y_i\frac{\partial }{\partial x_i}\bigg),
\end{equation}
\begin{equation}
\lk = -i \biggl( k_x\frac{\partial }{\partial k_y}-k_y\frac{\partial}{\partial k_x}\biggr)
\label{lk}
\end{equation}
are the OAM operators in real and momentum space.

Converting $a_{\vk }^{(r)}$ and $a_{\vk }^{(r)\dagger }$ to the Boson operators $b_{\vk }^{(n)}$ and $b_{\vk }^{(n)\dagger }$ 
that diagonalize the Hamiltonian in terms of the eigenmodes $\vert \vk,n\rangle = b_{\vk }^{(n)\dagger }\vert 0\rangle $ with frequencies $\omega_n(\vk)$ ($n=1,2$), 
we define \cite{fishmanbook18}
\begin{eqnarray}
a_{\vk }^{(r)}&=&\sum_{n=1}^2 \Bigl\{ X^{-1}(\vk )_{rn}\,b_{\vk}^{(n)} +X^{-1}(\vk )_{r,n+2}\,b_{-\vk}^{(n)\dagger }\Bigr\} , \\
a_{-\vk }^{(r)\dagger }&=&\sum_{n=1}^2 \Bigl\{ X^{-1}(\vk )_{r+2,n}\,b_{\vk}^{(n)} +X^{-1}(\vk )_{r+2,n+2}\,b_{-\vk}^{(n)\dagger }\Bigr\} .
\nonumber
\end{eqnarray}
The zero-temperature expectation value of ${\cal L}_z$ in eigenmode $n$
is then given by
\begin{eqnarray}
{\cal L}_{zn}(\vk )&=&\langle \vk ,n \vert {\cal L}_z \vert \vk ,n \rangle \nonumber \\
&=& \frac{\hbar}{2}\sum_{r=1}^2\Bigl\{ 
X^{-1}(\vk )_{rn}\,\lk \, X^{-1}(\vk )_{rn}^* \nonumber \\
&&- X^{-1}(\vk)_{r+2,n}\,\lk \, X^{-1}(\vk )^*_{r+2,n}\Bigr\}.
\label{Lzdef}
\end{eqnarray}
For collinear spin states {\it without} symmetry-breaking DM interactions, $\underline{X}^{-1}(-\vk )=\underline{X}^{-1}(\vk )^*$ so that
${\cal L}_{zn }(\vk )=-{\cal L}_{zn }(-\vk )$ is an odd function of $\vk $.  
The $4\times 4$-dimensional matrix $\underline{X}(\vk )$ will feature prominently in our subsequent discussion.

Due to the linear terms $k_x$ and $k_y$ in the OAM operator $\lk $ of Eq.\,(\ref{lk}), the resulting
${\cal L}_{zn}(\vk )$ is {\it not} a periodic function of $\vk $ in reciprocal space.
This shortcoming arises because the ``pseudo-momentum" $\vp_i$ in ${\cal L}_z =\sum_i (\vR_i\times \vp_i)\cdot \vz$ 
contains the derivatives $\partial /\partial x_i$ and $\partial /\partial y_i$ from Eq.\,(\ref{mom}) 
that enter the real-space OAM operator $\hat{l}_{zi}$ of Eq.\,(\ref{lzr}).
On a discrete lattice, those continuous derivatives should be replaced by finite differences.

\begin{figure}
\begin{center}
\includegraphics[width=6cm]{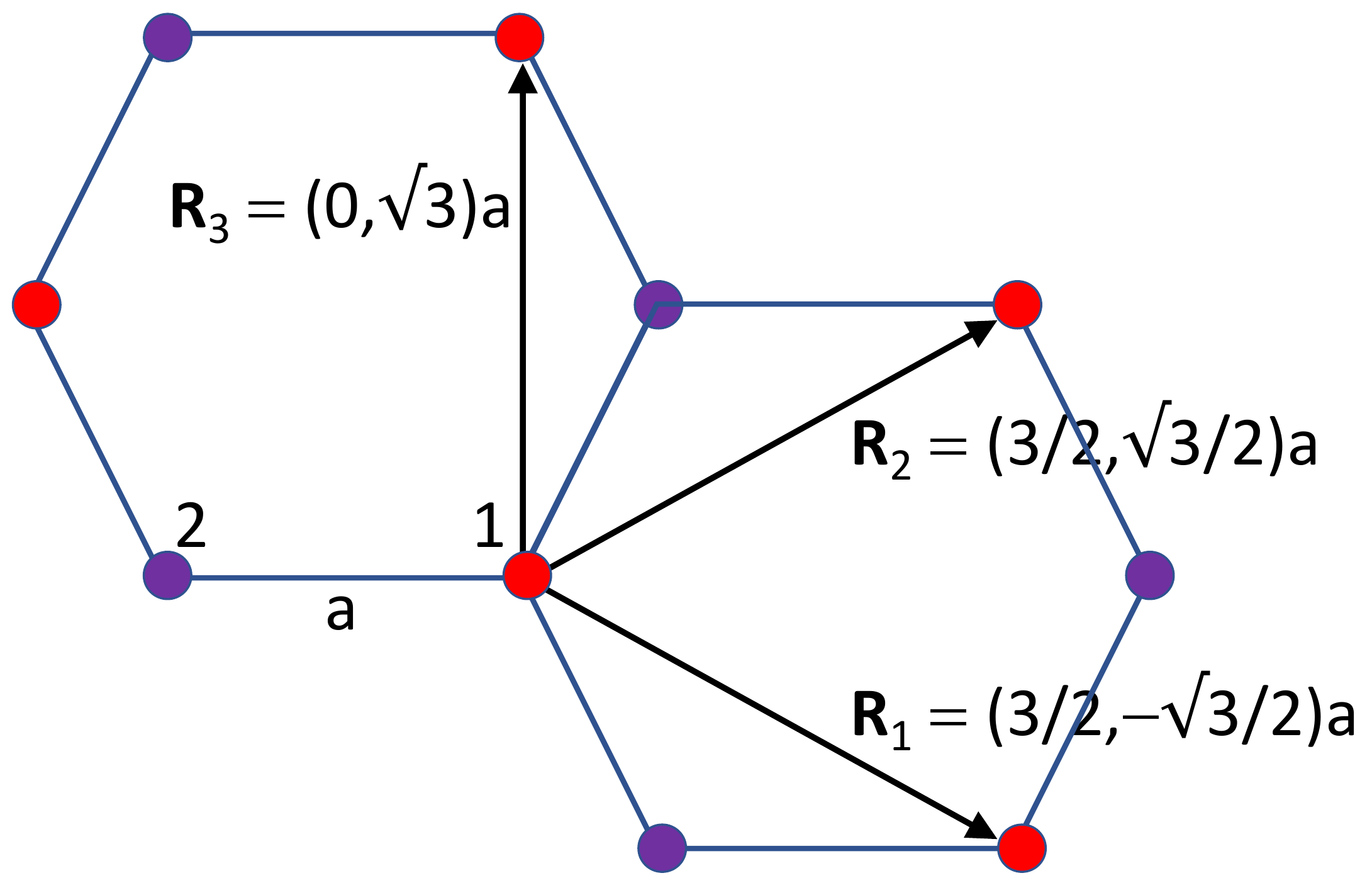}
\end{center}
\caption{A honeycomb lattice with unique sites 1 and 2 and 
the translation vectors ${\bf R}_1 = (3/2,-\sqrt{3}/2)a$, ${\bf R}_2 = (3/2,\sqrt{3}/2)a$,
and ${\bf R}_3 = (0,\sqrt{3})a$ coupling neighboring sites 1. }
\label{Fig1}
\end{figure}

The finite difference of a discrete function $f(\vr)$ produced by the translation vector $\vR_l$ is 
\begin{eqnarray}
\delta_l f(\vr) = \frac{1}{2 |\vR_l|} \Bigl\{ f(\vr + {\bf R}_l) - f(\vr - {\bf R}_l) \Bigr\}.
\end{eqnarray}
Transforming the continuous derivative $\partial f(\vr )/\partial r_\alpha$ 
by summing over distinct translation vectors $\vR_l$, we obtain
\begin{eqnarray}
\Delta_\alpha f(\vr) &=&\sum_l \frac{R_{l\alpha }}{\vert {\bf R}_l \vert }\, \delta_l f(\vr ) \nonumber \\
&=& \sum_l
\frac{R_{l \alpha}}{2 |{\bf R}_l|^2} \Bigl\{ f(\vr + {\bf R}_l) - f(\vr - {\bf R}_l) \Bigr\} ,
\end{eqnarray}
where $R_{l \alpha}$ is the projection of ${\bf R}_l$ along the $\alpha$ axis.
Note that $\Delta_{\alpha }f(\vr )\rightarrow \partial f(\vr )/\partial r_\alpha$ 
when the lattice translation vectors $\vR_l$ are orthogonal and their sizes vanish.
 
The Fourier transform of a magnon annihilation operator contains the factor $\exp ( i \vk \cdot \vr) $, which has
the discrete difference
\begin{eqnarray}
\Delta_\alpha \exp ( i \vk \cdot \vr) &=& i \exp ( i \vk \cdot \vr)\sum_l \frac{R_{l \alpha}}{|{\bf R}_l|^2} \sin (\vk \cdot {\bf R}_l)\nonumber 
\\
&=&i\kov_{\alpha }\exp ( i \vk \cdot \vr)
\end{eqnarray}
with periodic function 
\begin{equation}
\kov_{\alpha } = \sum_l \frac{R_{l\alpha}}{|{\bf R}_l|^2} \sin (\vk \cdot {\bf R}_l).
\end{equation}
For the honeycomb lattice in Fig.\,1, summing over the three translation vectors $\vR_l$ that couple site 1 to three
neighboring sites of type 1 produces
\begin{eqnarray}
\kov_xa&=&\sin(3k_xa/2)\cos(\sqrt{3}k_ya/2),\nonumber \\
\kov_ya&=& \frac{1}{\sqrt{3}}\Bigl\{ \sin(\sqrt{3}k_ya/2)\cos(3k_xa/2) \nonumber \\
&&+\sin(\sqrt{3}k_ya)\Bigr\}.
\end{eqnarray}
In the limit of small $k_x$ and $k_y$, $\kov_x\rightarrow 3k_x/2$ and $\kov_y\rightarrow 3k_y/2$.

The revised OAM operator in momentum space is then given by
\begin{equation}
\lk = -i \biggl( \kov_x\frac{\partial }{\partial k_y}-\kov_y\frac{\partial}{\partial k_x}\biggr).
\label{lka}
\end{equation}
In the limit $a\rightarrow 0$, there is no need to replace the {\it momentum space} derivatives 
$\partial /\partial k_x$ and $\partial /\partial k_y$ by finite differences.
Using the periodic functions $\kov_x$ and
$\kov_y$ instead of $k_x$ and $k_y$ imposes a natural bound on the OAM.
As seen below, this replacement has important consequences for the OAM at the high-symmetry points of the honeycomb lattice.

For a FM state in the limit of small energies and momenta, our results for the OAM (Eqs.\,(\ref{Lzdef}) and (\ref{lka})) reduce to the 
expression of Matsumoto and Murakami \cite{Mat11a, Mat11b}, 
which was parameterized in terms of a density-of-states and an effective mass.  Since we are interested in the OAM of both FM and AF magnons 
throughout the BZ, we prefer to work with the more general expressions given above.

\begin{figure}
\begin{center}
\includegraphics[width=6cm]{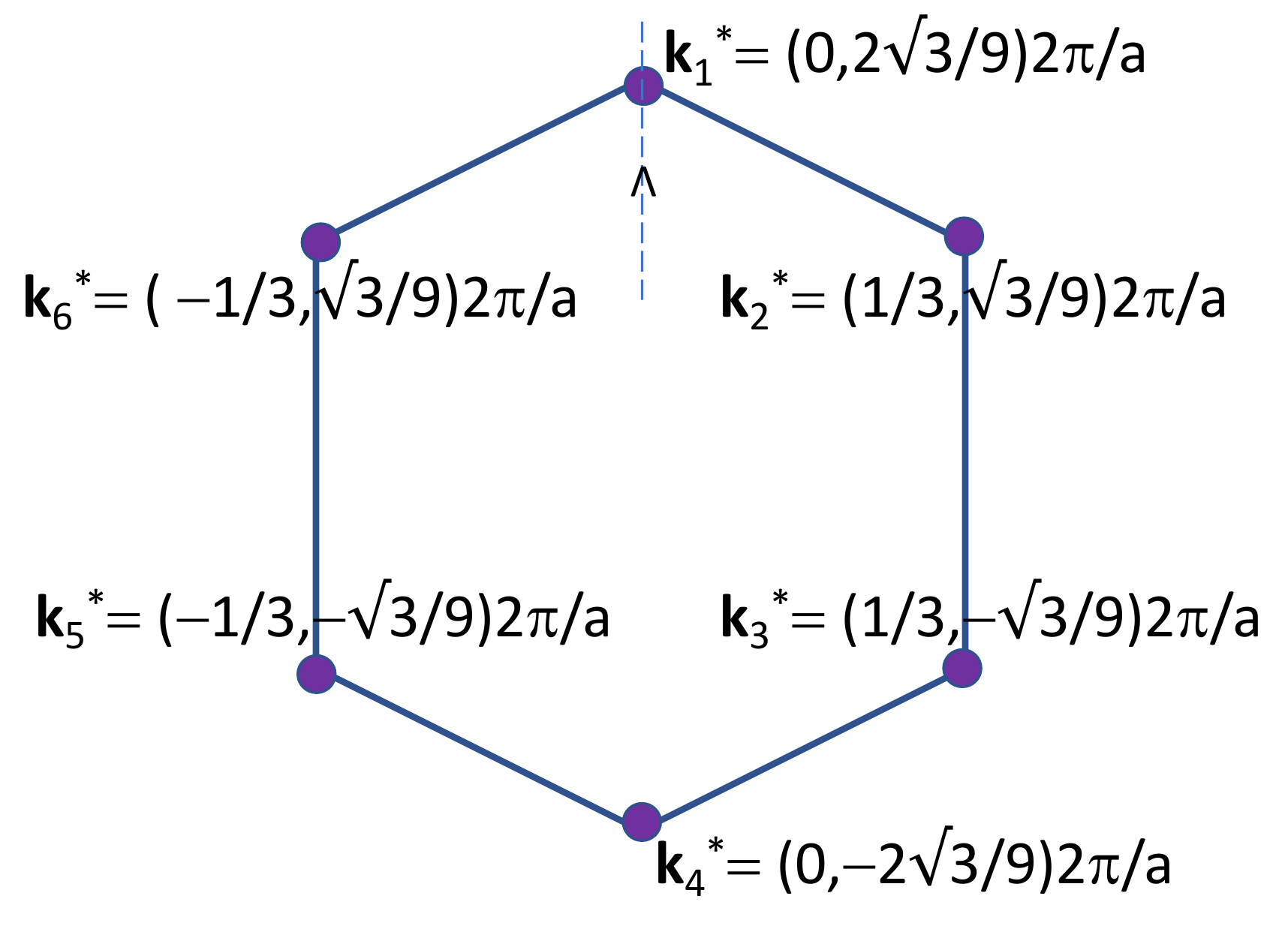}
\end{center}
\caption{The reciprocal-space hexagon of the honeycomb lattice showing the high-symmetry points $\vk^*$.  Direction
of approach to $\vk_1^*$ used to evaluate the OAM is drawn as an arrow.}
\label{Fig2}
\end{figure}

\section{FM Honeycomb with $D=0$}

We first consider the FM honeycomb lattice shown in Fig.\,3(a) with spins up, exchange coupling $J>0$, and no DM interaction.
The Hamiltonian is then
\begin{equation}
H=-J\sum_{\langle i,j\rangle} \vS_i \cdot  \vS_j ,
\end{equation}
where the sum runs over all nearest neighbors of the honeycomb lattice. 
Second order in the vector operator ${\bf v}_{\vk } =(a_{\vk }^{(1)},a_{\vk }^{(2)},a_{-\vk}^{(1)\dagger },a_{-\vk }^{(2)\dagger })$,
the Hamiltonian $H_2=\sum_\vk {\bf v}_{\vk}^{\dagger }\cdot \underline{{\it L}}(\vk )\cdot {\bf v}_{\vk }$ is given by
\begin{equation}
\underline{L}(\vk ) =
\frac{3JS}{2} \left(
\begin{array}{cccc}
1  & -\Gamma_{\vk}^* & 0 & 0 \\
-\Gamma_{\vk } & 1  & 0 & 0 \\
0 & 0 & 1  & - \Gamma_{\vk }^*\\
0 & 0 &-\Gamma_{\vk } & 1 \\
\end{array} \right),
\end{equation}
where 
\begin{eqnarray}
\Gamma_{\vk} &=&\frac{1}{3}\Bigl\{ e^{ik_xa}+e^{-i(k_x+\sqrt{3}k_y)a/2}\nonumber \\
&+&e^{-i(k_x-\sqrt{3}k_y)a/2}\Bigr\}.
\end{eqnarray}
The magnon dynamics is determined by diagonalizing $\underline{L}\cdot \underline{N}$, where
\begin{equation}
\underline{N} =
\left(
\begin{array}{cc}
\underline{I} &0 \\
0 & -\underline{I} \\
\end{array} \right)
\end{equation}
and $\underline{I}$ is the two-dimensional identity matrix.
Using the relation $\underline{N}\cdot \underline{X}^{\dagger }(\vk )\cdot \underline{N}=\underline{X}^{-1}(\vk)$ to normalize the 
eigenvectors \cite{fishmanbook18} $X^{-1}(\vk )_{rn}$, we find
\begin{equation}
\underline{X}^{-1}(\vk ) =
\frac{1}{\sqrt{2}\,\Gamma_{\vk }^*}
\left(
\begin{array}{cccc}
-\Gamma_{\vk}^* & \Gamma_{\vk }^* & 0 & 0 \\
\vert \Gamma_{\vk }\vert  & \vert \Gamma_{\vk }\vert & 0 & 0 \\
0 & 0 & -\Gamma_{\vk }^* & \Gamma_{\vk }^*\\
0 & 0 &\vert \Gamma_{\vk }\vert  & \vert \Gamma_{\vk }\vert  \\
\end{array} \right).
\label{xin1}
\end{equation}
It is then simple to show that 
\begin{equation}
\label{lz1}
{\cal L}_{zn}(\vk )=\frac{\hbar}{4}\frac{\Gamma_{\vk }}{\vert \Gamma_{\vk }\vert }\,\lk \,
\frac{\Gamma_{\vk }^*}{\vert \Gamma_{\vk }\vert }
\end{equation}
is the same for magnon bands $n=1$ and 2 with frequencies $\omega_{1,2}(\vk )=3JS(1\pm \vert \Gamma_{\vk }\vert )$
(1 for +, 2 for $-$).

The upper and lower mode frequencies $\omega_1(\vk )$ and $\omega_2(\vk )$ cross at wavevectors $\kkon =(0,2\sqrt{3}/9)2\pi/a$, $\kktw =(1/3,$ $\sqrt{3}/9)2\pi/a$, 
and symmetry-related points $\vk^*$ at the corners of the BZ where $\Gamma_{\vk^* }=0$.  
As seen in Fig.\,3(b), ${\cal L}_{zn}(\vk^* )$ has alternating values of $\sim \pm 0.185\hbar $ at these
crossing points with the positive value $\sim 0.185 \hbar $ at $\kkon $,

To analytically evaluate the OAM at $\vk_1^*$, we construct the unit vector ($\vk \ne \vk^*_1$)
\begin{eqnarray}
{\hat u}(\vk )&\equiv & \frac{\Gamma_{\vk }}{\vert \Gamma_{\vk }\vert} = \frac{e^{ik_xa/4}}{C(\vk )} \Bigl\{ A_+(\vk ) \cos (3k_xa/4) \nonumber \\
&+& iA_-(\vk )\sin(3k_xa/4 )\Bigr\},
\end{eqnarray}
where
$A_\pm (\vk )=1\pm 2\cos (\sqrt{3}k_ya/2)$ and
\begin{eqnarray}
C(\vk ) &=& \Bigl\{ 1 + 4\cos^2 (\sqrt{3}k_ya/2)\nonumber \\
&+&4\cos (\sqrt{3}k_ya/2)\cos (3k_xa/2)\Bigr\}^{1/2}.
\end{eqnarray}
Because
\begin{equation}
{\cal L}_{zn}(\vk )=-\frac{i\hbar }{4}{\hat u}(\vk ) \biggl( \kov_x\frac{\partial {\hat u}(\vk )^*}{\partial k_y}-\kov_y\frac{\partial {\hat u}(\vk )^* }{\partial k_x}\biggr) 
\end{equation}
is singular at $\vk^*_1$, we calculate the OAM at fixed $k^*_{1x}=0$ 
while $k_{1y}^*=(2\sqrt{3}/9)2\pi/a$ is approached from below, as sketched 
in Fig.\,2.  (Of course, the OAM can also be evaluated by approaching $\vk_1^*$ from above or from
any other direction.)   We shall then evaluate the OAM at other BZ corners $\vk^*_m$ by applying a rotation operator.

\begin{figure}
\begin{center}
\includegraphics[width=8cm]{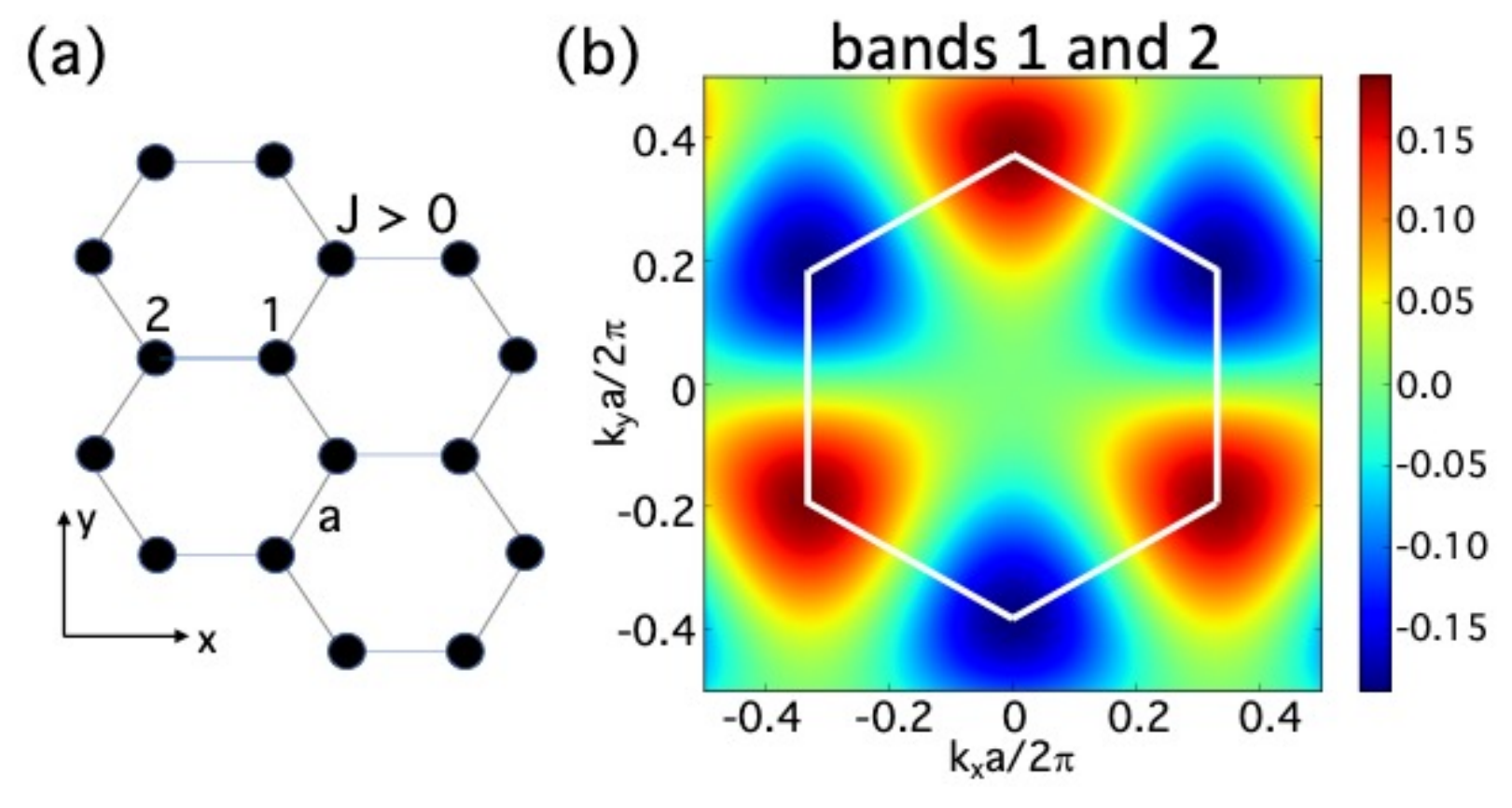}
\end{center}
\caption{(a)  A honeycomb lattice with FM exchange $J>0$.  (b) The OAM for the bands 1 and 2 versus $\vk $.}
\label{Fig3}
\end{figure}

When $k_x=0$ but $k_y < (2\sqrt{3}/9 )2\pi/a$, 
$\Gamma_{\vk }=1+2\cos (\sqrt{3}k_ya /2)$ and ${\hat u}(\vk )= 1$.
It is then straightforward to show that
\begin{equation}
\lim_{k_x \rightarrow 0} \frac{\partial {\hat u}(\vk )^* }{\partial k_x} = -\frac{3ia}{2} \frac{1}{1+2\cos (\sqrt{3}k_ya/2)}.
\end{equation}
Since $\bar{k}_x\,\partial {\hat u}(\vk )/\partial k_y$ vanishes as $k_x \rightarrow 0$, we obtain
\begin{eqnarray}
\label{lzlim}
{\cal L}_{zn}(\vk^*_1)&=& \frac{i\hbar }{4} \lim_{\vk \rightarrow \vk^*_1}  \bar{k}_y \frac{\partial {\hat u}(\vk )^*}{\partial k_x}\nonumber \\
&=& \frac{\sqrt{3}\hbar }{8} \lim_{k_y \rightarrow 4\pi\sqrt{3}/9a} \frac{\sin(\sqrt{3}k_ya/2)+\sin(\sqrt{3}k_ya)}{1+2\cos (\sqrt{3}k_ya/2)} \nonumber \\
& =& \frac{3\hbar }{16}= 0.1875 \hbar ,
\end{eqnarray}
which uses $\bar{k}_ya=\big(\sin(\sqrt{3}k_ya/2)+\sin(\sqrt{3}k_ya)\big)/\sqrt{3}$.
Notice that it is essential to use the periodic function $\bar{k}_y$ in the OAM operator $\hat{l}_{z\vk }$
and in Eq.\,(\ref{lzlim}) to 
produce a finite result for ${\cal L}_{zn}(\vk^*_1)$.

Suppose wavevector ${\vk}' =\underline{R}_z(\theta )\cdot \vk $ is obtained from wavevector $\vk $
by applying the rotation operator 
\begin{equation}
\underline{R}_z(\theta ) =
\left(
\begin{array}{cc}
\cos \theta  & \sin \theta \\
-\sin \theta & \cos \theta  \\
\end{array} \right)
\end{equation}
about the $z$ axis.
For $\theta =\pi /3$, it is easy to show that $\Gamma_{\vk'}=\Gamma_{\vk }^*=\Gamma_{-\vk }$ and, using Eq.\,(\ref{xin1}), 
that $\underline{X}^{-1}(\vk')=\underline{X}(\vk )^*
=\underline{X}(-\vk )$.  So with high-symmetry points $\vk^*_m=\underline{R}_z(\pi/3)\cdot \vk^*_{m-1}$ and $\vk^*_{m-1}$ 
rotated with respect to one another by $\pi /3$,
${\cal L}_{zn}(\vk^*_m)={\cal L}_{zn}(-\vk^*_{m-1})
=-{\cal L}_{zn}(\vk^*_{m-1})$.  Hence, the OAM changes sign 
around the reciprocal-space hexagon in Fig.\,2.  

\section{FM Honeycomb with DM interaction}

We next consider the FM honeycomb lattice shown in the inset to Fig.\,4 with spins up and DM interaction $D$ between next-nearest neighbors.
The Hamiltonian is then given by
\begin{eqnarray}
H&=&-J\sum_{\langle i,j\rangle} \vS_i \cdot  \vS_j -D \sum_{\langle \langle i,j\rangle \rangle }' \vz\cdot (\vS_i \times \vS_j) \nonumber \\
&+&D\sum_{\langle \langle i,j\rangle \rangle }^{''} \vz \cdot (\vS_i \times \vS_j) -K\sum_i S_{iz}^2 ,
\end{eqnarray}
where the first DM sum runs over the first triangle shown in the inset to Fig.\,4 and the second sum with opposite sign runs over the second triangle.
The anisotropy $K >0$ is assumed to be sufficiently large to prevent the spins from tilting away from the $z$ axis.

\begin{figure}
\begin{center}
\includegraphics[width=8cm]{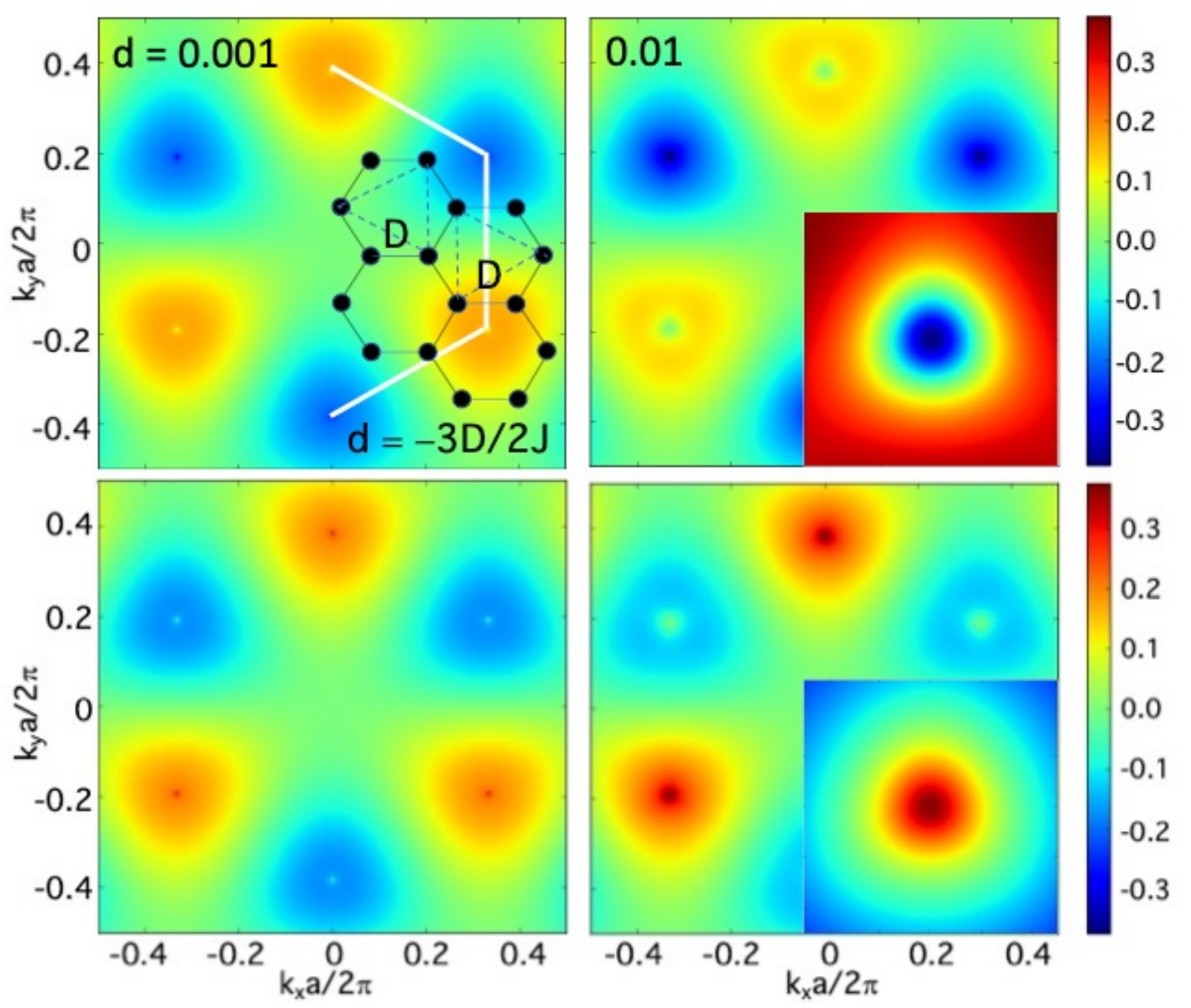}
\end{center}
\caption{
The OAM of the FM honeycomb lattice with DM interaction $d=-3D/2J$ versus $\vk $.  Results for the upper band are shown at the top and for the
lower band at the bottom.  For $d=0.01$, insets blow up the regions around $\vk_1^*$ with widths $\Delta k_{x,y} = 0.2\pi /a$}
\label{Fig4}
\end{figure}

We then obtain
\begin{equation}
\underline{L}(\vk ) =
\frac{3JS}{2} \left(
\begin{array}{cccc}
1-G_{\vk }  & -\Gamma_{\vk}^* & 0 & 0 \\
-\Gamma_{\vk } & 1+G_{\vk }  & 0 & 0 \\
0 & 0 & 1+G_{\vk}  & - \Gamma_{\vk }^*\\
0 & 0 &-\Gamma_{\vk } & 1-G_{\vk} \\
\end{array} \right),
\end{equation}
where $G_{\vk }=d \,\Theta_{\vk }$ with $d = -2D/3J$ and
\begin{equation}
\label{Th}
\Theta_{\vk} = 4\cos(3k_xa/2) \sin(\sqrt{3} k_ya/2)-2\sin(\sqrt{3}k_ya).
\end{equation} 
Because the anisotropy $\kappa =2K/3\vert J\vert $ merely shifts the magnon energies 
\begin{equation}
\hbar \omega_{1,2}(\vk)=3JS \Bigl\{ 1 + \kappa \pm \sqrt{\vert \Gamma_{\vk }\vert^2 +G_{\vk }^2}\Bigr\}
\end{equation}
but does not affect the OAM,
we neglect its contribution to $\underline{L}(\vk )$.

It follows that 
\begin{equation}
\underline{X}^{-1}(\vk ) =
\frac{1}{\sqrt{2}\,\Gamma_{\vk }^*}
\left(
\begin{array}{cccc}
-\Gamma_{\vk }^* F_{\vk }^-& \Gamma_{\vk }^* F_{\vk }^+ & 0 & 0 \\
\vert \Gamma_{\vk }\vert  F_{\vk }^+ & \vert \Gamma_{\vk }\vert F_{\vk }^- & 0 & 0 \\
0 & 0 & -\Gamma_{\vk }^* F_{\vk }^- & \Gamma_{\vk }^* F_{\vk }^+\\
0 & 0 &\vert \Gamma_{\vk }\vert  F_{\vk }^+ & \vert \Gamma_{\vk }\vert F_{\vk }^- \\
\end{array} \right),
\label{xin1d}
\end{equation}
where
\begin{equation}
F_{\vk}^{\pm }= 1 \pm \frac{G_{\vk }}{\sqrt{\vert \Gamma_{\vk }\vert^2 + G_{\vk }^2}}.
\end{equation}
Since $F_{\vk }^{\pm }$ is real, it is simple to show that 
\begin{equation}
\label{lz1d}
{\cal L}_{z1}(\vk )=\frac{\hbar}{4}F_{\vk }^+ \frac{\Gamma_{\vk }}{\vert \Gamma_{\vk }\vert }\,\lk \,
\frac{\Gamma_{\vk }^*}{\vert \Gamma_{\vk }\vert },
\end{equation}
\begin{equation}
\label{lz2d}
{\cal L}_{z2}(\vk )=\frac{\hbar}{4}F_{\vk }^- \frac{\Gamma_{\vk }}{\vert \Gamma_{\vk }\vert }\,\lk \,
\frac{\Gamma_{\vk }^*}{\vert \Gamma_{\vk }\vert }
\end{equation}
for the lower and upper magnon bands, respectively.
Because $G_{-\vk }=-G_{\vk }$ and $F_{-\vk}^{\pm }=F_{\vk }^{\mp }$, the OAM satisfies the relation
${\cal L}_{z1 }(\vk )=-{\cal L}_{z2 }(-\vk )$ and is no longer an odd function of $\vk $ within a single band.

Numerical results for the OAM at $d=0.001$ and 0.01 are plotted in Fig.\,4.  While the OAM for the lower
band on the bottom of Fig.\,4 is biased towards positive values, the OAM for the upper band on the top is biased towards negative values.  
Nevertheless, the sum of the OAM over the two bands vanishes, as guaranteed by the symmetry relation above.  

These 
results are verified by the one-dimensional slice along the edge of the BZ plotted in Fig.\,5 with $k_xa/2\pi =1/3$ and
$k_ya/2\pi $ running from $-\sqrt{3}/9$ to $\sqrt{3}/9$.   The corners $\vk_3^*$ and $\vk_2^*$ of the BZ lie at the beginning and end of this path.
For $d=0$, the OAM equals $\pm 0.1875\hbar $ at the corners of the BZ for both bands.  But even a small DM interaction
changes that result dramatically.   

\begin{figure}
\begin{center}
\includegraphics[width=8cm]{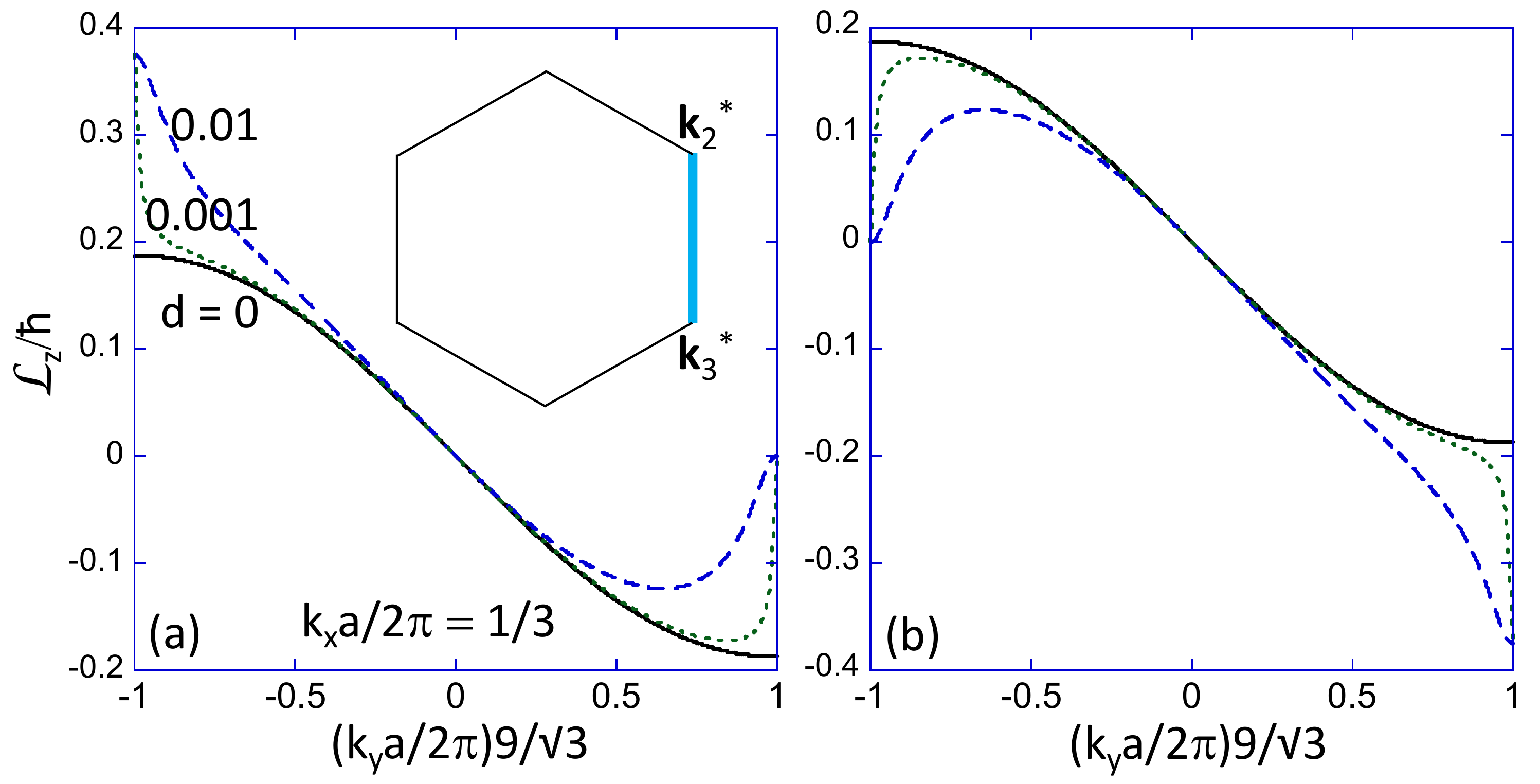}
\end{center}
\caption{
The OAM of the FM honeycomb lattice along the vertical edge of the BZ with $k_x a/2\pi =1/3$ from $k_ya/2\pi =-\sqrt{3}/9$ to $\sqrt{3}/9$ with 
$d=0$, 0.001, and 0.01.  Results for the lower and upper bands are shown in (a) and (b), respectively.}
\label{Fig4}
\end{figure}

This can be readily seen from an exact solution.
Since $\Theta_{\vk_{2m}^*}=3\sqrt{3}>0$ and $\Theta_{\vk_{2m+1}^*}=-3\sqrt{3}<0$, it follows that $F_{\vk_{2m}^*}^+ = 2$, $F_{\vk_{2m+1}^*}^+ = 0$,
$F_{\vk_{2m}^* }^-=0$, and $F_{\vk_{2m+1}^*}^- =2$.
Using the results of the previous section, we then find that for $d\ne 0$,
\begin{eqnarray}
\label{lzlimd}
{\cal L}_{z1}(\vk^*_{2m})&=&\frac{3\hbar }{8}= 0.375 \hbar ,\\
{\cal L}_{z1}(\vk^*_{2m+1})&=&0,\\
{\cal L}_{z2}(\vk^*_{2m})&=&0,\\
{\cal L}_{z2}(\vk^*_{2m+1})&=&-\frac{3\hbar }{8}= -0.375 \hbar ,
\end{eqnarray}
in agreement with the numerical results of Fig.\ 5.
Remarkably, these exact results are satisfied in the presence of infinitesimal values of the DM interaction.  But  
as can be easily seen from Figs.\,4 and 5, the region around $\vk^*$ with a large or vanishingly small value of ${\cal L}_{zn}(\vk )$ 
shrinks as $d\rightarrow 0$.
Notice that Figs.\,5(a) and (b) satisfy the symmetry relation ${\cal L}_{z1 }(\vk )=-{\cal L}_{z2 }(-\vk )$.

\section {AF Honeycomb}

Finally, we consider the AF honeycomb lattice in Fig.\,6(a) with alternating up and down spins, exchange coupling 
$J<0$, easy-axis anisotropy $K$, and DM interaction $D$. Then 
\begin{equation}
\underline{L}(\vk ) =
-\frac{3JS}{2} \left(
\begin{array}{cccc}
1+\kappa  & 0 & 0 & -\Gamma_{\vk}^* \\
0 & 1+\kappa    &  -\Gamma_{\vk} & 0\\
0 & -\Gamma_{\vk}^* & 1+\kappa  & 0\\
-\Gamma_{\vk} & 0 &0 & 1+\kappa  \\
\end{array} \right)
\end{equation}
with corresponding
\begin{equation}
\underline{X}^{-1}(\vk ) =
\frac{1}{\sqrt{2f_{\vk }}\,\Gamma_{\vk }^*}
\left(
\begin{array}{cccc}
-h^+_{\vk } \Gamma_{\vk }^*& 0 & 0 & h^-_{\vk }\Gamma_{\vk }^* \\
0  &  h^-_{\vk } g_{\vk }^+ & -h^+_{\vk } g_{\vk }^-  & 0 \\
0 & h^-_{\vk }\Gamma_{\vk }^* & -h^+_{\vk }\Gamma_{\vk }^* & 0 \\
-h^+_{\vk } g_{\vk }^- & 0 & 0 & h_{\vk }^-g_{\vk}^+    \\
\end{array} \right),
\end{equation}
where $f_{\vk } =\sqrt{(1+\kappa)^2-\vert \Gamma_{\vk }\vert^2}$, $g_{\vk }^{\pm }=1+\kappa \pm f_{\vk }$,
and $h_{\vk }^{\pm }=\sqrt{g_{\vk }^{\pm }}$.

\begin{figure}
\begin{center}
\includegraphics[width=8cm]{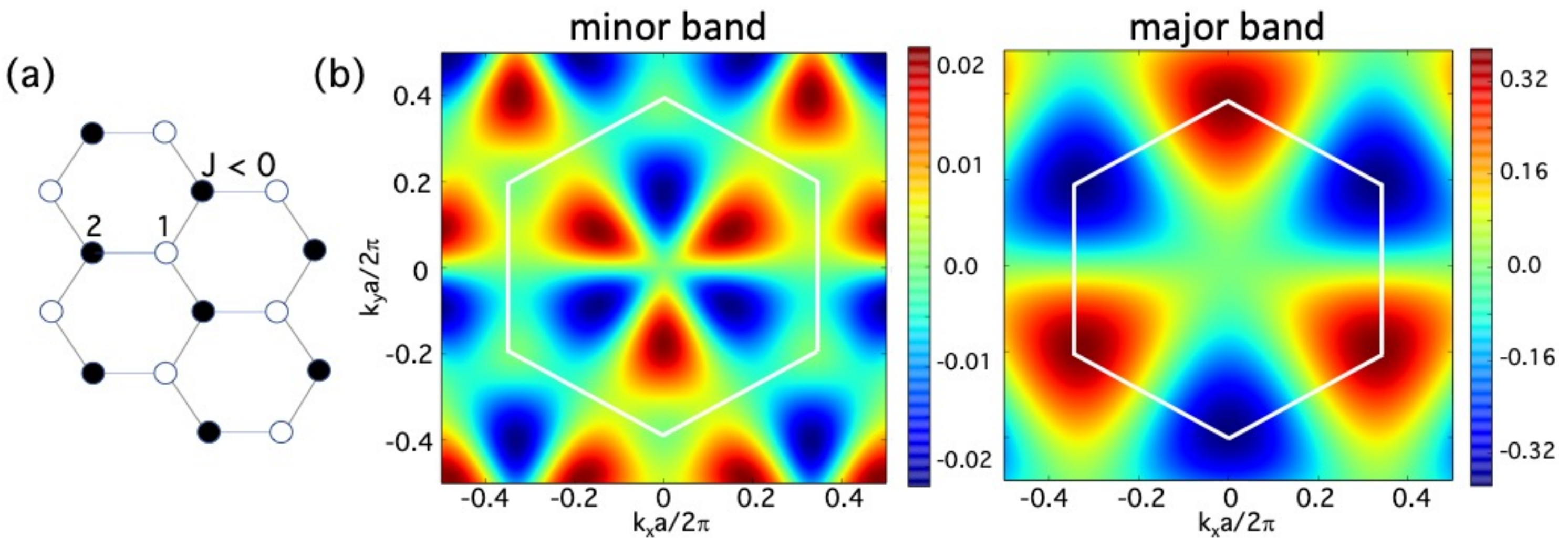}
\end{center}
\caption{(a) A honeycomb lattice with AF exchange $J<0$ between up (closed circles, site 2) and down (open circles, site 1) spins.
(b) The OAM of the minor (left) and major (right) bands versus $\vk $ for $\kappa =0$.}
\label{Fig6}
\end{figure}

Because the DM interactions shift the degenerate mode frequencies
\begin{equation}
\omega_{1,2}(\vk )=3\vert J\vert S\Bigl\{ \sqrt{(1+\kappa )^2-\vert \Gamma_{\vk }\vert^2}+ G_{\vk } \Bigr\}
\end{equation}
through the $G(\vk )$ term but do not affect the OAM, we neglect their contribution to $\underline{L}(\vk )$ above.
Surprisingly, the degenerate magnon bands exhibit distinct OAM with 
\begin {eqnarray}
\label{lzhc1}
{\cal L}_{z1}(\vk )&=&\frac{\hbar }{4}\frac{1 +\kappa +f_{\vk }}{f_{\vk }} \,
\frac{\Gamma_{\vk} }{\vert \Gamma_{\vk }\vert }\,\lk \, \frac{\Gamma_{\vk }^* }{\vert \Gamma_{\vk }\vert },\\
\label{lzhc2}
{\cal L}_{z2}(\vk )&=&-\frac{\hbar }{4}\frac{1 +\kappa -f_{\vk }}{f_{\vk }}\,
\frac{\Gamma_{\vk} }{\vert \Gamma_{\vk }\vert }\,\lk \, \frac{\Gamma_{\vk }^* }{\vert \Gamma_{\vk }\vert }.
\end{eqnarray}
As seen in Fig.\,6(b), the major ($n=1$) and minor ($n=2$) bands have different OAM patterns 
but are both threefold symmetric.  
Notice that ${\cal L}_{{\rm av }}(\vk )= ({\cal L}_{z1}(\vk )+{\cal L}_{z2}(\vk ))/2$ for the two
bands of the AF honeycomb lattice equals the OAM ${\cal L}_{zn}(\vk )$ of the FM honeycomb lattice with $D=0$, given by Eq.\,(\ref{lz1}) and plotted in Fig.\,3(b). 

At the BZ corners $\vk^*$, $\Gamma_{\vk^* }=0$, $f_{\vk^* }=1$, and the magnon frequencies reach maxima of $\omega_n(\vk^*)=3\vert J\vert S(1+\kappa \pm 3\sqrt{3}d)$
at $\vk_{2m}^*$ and $\vk_{2m+1}^*$, respectively. 
Comparing Eqs.\,(\ref{lzhc1}) and ({\ref{lzhc2}) with Eq.\,(\ref{lz1}), we find that ${\cal L}_{z1}(\vk^*)=\pm (3\hbar /8 )(1+\kappa /2)$ and ${\cal L}_{z2}(\vk^*)
=\mp (3\hbar /16)\kappa $.
Hence, the average amplitude $\vert {\cal L}_{{\rm av }}(\vk^*)\vert $ of the two bands of the AF honeycomb lattice equals $3\hbar /16$, 
as expected.

\section{Conclusion}

The theory developed in this paper can be easily extended to treat many real three-dimensional materials that contain
honeycomb lattices lying within their $ab$ planes.  Sivadas {\it et al.} \cite{Sivadas15} reviewed the 
the magnetic phase diagrams of honeycomb systems with chemical formula ABX$_3$.
Some examples of FM honeycomb lattices within that class are CrSiTe$_3$ and CrGeTe$_3$ \cite{Carteaux95a, Carteaux95b, Casto15}.  
Two other Cr-based FM honeycomb systems are CrI$_3$ \cite{McGuire15, Chen18t} and
CrCl$_3$ \cite{Li22}.  Two examples of 
AF honeycomb lattices are MnPS$_3$ and MnPSe$_3$ \cite{Wildes98}.  
 
The exact calculations in this paper reveal several pertinent points about the OAM of magnons on honeycomb lattices.  First, finite results for the OAM at high-symmetry 
points $\vk^*$ are only reached if periodic functions $\kov_x$ and $\kov_y$ are used in the OAM operator $\lk $.  If non-periodic terms $k_x$ and $k_y$ were retained in the OAM 
operator, then
the OAM would diverge at the high-symmetry points $\vk^*$.  Second, the OAM of the FM and AF honeycomb lattices are closely connected.  The average OAM of the major and minor AF honeycomb bands
equals the OAM of the FM bands.  Third, the largest OAM for the pure honeycomb lattice with amplitude $3\hbar /8$ is still substantially less than $\hbar $.  So the spin angular momentum of 
a single spin flip is bigger than the largest OAM.  Fourth, 
the DM interaction produced by SO coupling controls the behavior of the OAM near the corners of the BZ, no matter how small the size of $d$.  We conclude that once 
the network of exchange interactions generates an OAM by confining the magnons, SO coupling can still have important effects.  Experimentally, the maximum OAM expected
for a FM honeycomb material with even a small to moderate value of the DM interaction should be about $0.375 \hbar $.

Although performed at zero temperature, these calculations indicate the available magnon states that can be filled at nonzero temperatures.  For the FM honeycomb lattice with $D=0$,
the symmetry relation ${\cal L}_{zn }(\vk )=-{\cal L}_{zn }(-\vk )$ for magnon bands $n=1$ and 2 implies that the thermal average $\langle {\cal L}_z\rangle $ vanishes due to the integral over wavevectors $\vk $.
As discussed above, the DM interaction $D$ breaks that symmetry relation leaving only ${\cal L}_{z1 }(\vk )=-{\cal L}_{z2 }(-\vk )$.  Consequently, the lower or upper magnon band will carry 
positive or negative OAM, respectively.  This implies that $\langle {\cal L}_z\rangle $ peaks at a temperature $T$ of order $D^2/J$.   We will examine this behavior more carefully in future work.

Even for the simple case of a honeycomb lattice with a single exchange interaction, nearest-neighbor DM interactions, and easy-axis anisotropy, 
many questions remain unanswered.  What is the coupling between the spin and orbital 
angular momentum:  do they tend to lie parallel or antiparallel?   What is the effect of next-neighbor exchange couplings for both FM and AF honeycomb lattices?  Of magnetic fields?  
Is there some easy way to understand the qualitative difference between the OAM of the major and minor bands of the AF honeycomb lattice?  Since the major and minor
bands lie at the same energy and momentum, can OAM be easily exchanged between the magnons of the two bands?
We hope that future research on the OAM of magnons in other non-Bravais lattices will be motivated by the exact results presented in this work.


We thank Jason Gardner for useful conversations.  Research sponsored by the U.S. Department of Energy, 
Office of Basic Energy Sciences, Materials Sciences and Engineering Division.
The data that support the findings of this study are available from the corresponding author
upon reasonable request.

\vfill

\end{document}